\documentclass[balance,upint,subscriptcorrection,varvw,mathalfa=cal=boondoxo,spanish,french,vietnamese,russian,greek,pdf-a,colorlinks]{asmeconf}

\usepackage{makecell}
\usepackage{algorithm}  
\usepackage{algpseudocode}
\usepackage{gensymb}  
\allowdisplaybreaks 

\hypersetup{%
	pdfauthor={Manuel Sage},									  
	pdftitle={The Economic Dispatch of Power-To-Gas Systems with Deep Reinforcement Learning: Tackling The Challenge of Delayed Rewards with Long-Term Energy Storage},                  
	pdfkeywords={economic dispatch, reinforcement learning},
	pdfsubject = {Conference paper},			  
}

\fancypagestyle{plain}{  
	\fancyhf{} 
	\fancyhead[L]{\textit{This paper is a preprint accepted for publication at ASME}} 
	\fancyfoot[C]{\thepage} 
}

\begin{document}




\title{The Economic Dispatch of Power-to-Gas Systems with Deep Reinforcement Learning:\\Tackling the Challenge of Delayed Rewards with Long-Term Energy Storage}
 
%
%
%

\SetAuthors{%
	Manuel Sage\affil{1}, 
	Khalil Al Handawi\affil{1}\affil{2}, 
	Yaoyao Fiona Zhao\affil{1}\CorrespondingAuthor{yaoyao.zhao@mcgill.ca}, 
	}

\SetAffiliation{1}{McGill University, Montreal, Canada}
\SetAffiliation{2}{Siemens Energy, Montreal, Canada}

\maketitle





\begin{abstract}
Power-to-Gas (P2G) technologies gain recognition for enabling the integration of intermittent renewables, such as wind and solar, into electricity grids. However, determining the most cost-effective operation of these systems is complex due to the volatile nature of renewable energy, electricity prices, and loads. Additionally, P2G systems are less efficient in converting and storing energy compared to battery energy storage systems (BESs), and the benefits of converting electricity into gas are not immediately apparent.
Deep Reinforcement Learning (DRL) has shown promise in managing the operation of energy systems amidst these uncertainties. Yet, DRL techniques face difficulties with the delayed reward characteristic of P2G system operation. Previous research has mostly focused on short-term studies that look at the energy conversion process, neglecting the long-term storage capabilities of P2G.

This study presents a new method by thoroughly examining how DRL can be applied to the economic operation of P2G systems, in combination with BESs and gas turbines, over extended periods. Through three progressively more complex case studies, we assess the performance of DRL algorithms, specifically Deep Q-Networks and Proximal Policy Optimization, and introduce modifications to enhance their effectiveness. These modifications include integrating forecasts, implementing penalties on the reward function, and applying strategic cost calculations, all aimed at addressing the issue of delayed rewards. Our findings indicate that while DRL initially struggles with the complex decision-making required for P2G system operation, the adjustments we propose significantly improve its capability to devise cost-effective operation strategies, thereby unlocking the potential for long-term energy storage in P2G technologies.

\end{abstract}


%
%



\section{Introduction}

Power-to-Gas (P2G) technologies are gaining attention from academia and industry alike due to the increasing integration of intermittent renewable energy sources into our power grids. P2G represents a range of methods for converting electrical power into gases such as hydrogen, ammonia, and synthetic natural gas (SNG). The underlying concept behind these technologies is to store energy for future use, addressing the challenge of the intermittency of renewable energy sources. This typically involves storing excess electricity that would otherwise be wasted or harnessing electricity when it's more affordable, for use during periods when renewable energy production is low, demand is high, and prices surge.

While battery energy storage systems (BESs) are generally better suited for short-term energy storage, P2G systems excel in the long-term storage of larger energy quantities. However, the efficiency of P2G systems in converting and storing energy is lower compared to BESs. Additionally, assessing the economic feasibility of P2G systems is challenging due to uncertainties in several critical factors, such as the variability of renewable energy production and electricity prices. Making informed decisions on when to directly use renewable energy, store it in BESs, or convert it to gas using P2G systems is complex.

Deep Reinforcement Learning (DRL) is emerging as a promising approach to tackle these uncertainties. By training decision-making on historical data, DRL can learn effective dispatch strategies for a wide range of possible scenarios. In recent years, numerous studies have applied DRL to dispatch problems for BESs, including for energy arbitrage, improved utilization of renewable energy, load following and combinations of these tasks \cite{sage2024_DRL_forecasting, sage2025_bes_comprehensive, a225_huang, a213_CAO_energyarbitrage, a237_HARROLD2022121958}.
Applications of DRL to P2G system dispatch are less frequent. Zhang et al. deployed the DRL models Deep Deterministic Policy Gradient (DDPG) and Deep Q-Networks (DQN) to learn dispatch strategies for a hybrid energy system with wind turbines, P2G system (producing SNG), and gas turbine (GT) \cite{a253_ZHANG2020113063}. A similar system, including renewables, P2G, GT, and BES, has successfully been controlled by DRL in the work of Teng et al. \cite{a179_teng2021}. The integrated electricity-natural gas system modeled by Liang et al. \cite{a258_LIANG2024123390} furthermore includes a carbon capture and storage facility that produces the CO2 required for methanation. Here, with DDPG, DQN, and Twin Delayed Deep Deterministic policy gradient (TD3), the authors compared three different DRL algorithms and found TD3 to perform best.

These existing studies utilize closely related problem formulations, focusing on the energy conversion aspect of P2G rather than the use as energy storage. With fluctuating renewables and a load profile for the electricity network, the P2G systems typically operate on power that would otherwise be curtailed. Furthermore, the GTs in the systems operate independently of the P2G systems due to connections to the gas grid. The durations of the case studies simulated in Refs. \cite{a253_ZHANG2020113063, a179_teng2021, a258_LIANG2024123390} are limited to 24 hours, which demonstrates that learning long-term dispatch strategies for the P2G systems was not the objective of these studies.

In this study, we take an alternative approach and test the ability of DRL models to learn long-term dispatch strategies for hybrid energy systems with P2G technology. For this purpose, we simulate a plant with wind turbines, BES, P2G through electrolyzation and methanation, and a GT that is connected to the power grid but not to the gas grid. Thus, only SNG previously generated by the P2G system can be burnt for power generation in the GT. The simulated plant participates in energy arbitrage and sells the power it produces to the utility grid without a limiting load. This removes excess renewable power from the equation and requires DRL agents to better understand the economics of P2G operation. The contributions of this study are as follows:
\begin{enumerate}
	\item We propose a novel hybrid energy system formulation including BES, P2G system, GT, and renewable energy that, other than existing studies, focuses on the long-term connections between P2G and GT dispatch.
	\item On three case studies with increasing complexity, with DQN and Proximal Policy Optimization (PPO), we adopt two popular DRL algorithms to our framework.
	\item We propose a number of problem formulation strategies, mainly modifications to the reward function, which successfully improve DRL performance on P2G system dispatch. 
\end{enumerate}


\begin{figure*}
	\centering
		\includegraphics[width=0.85\textwidth]{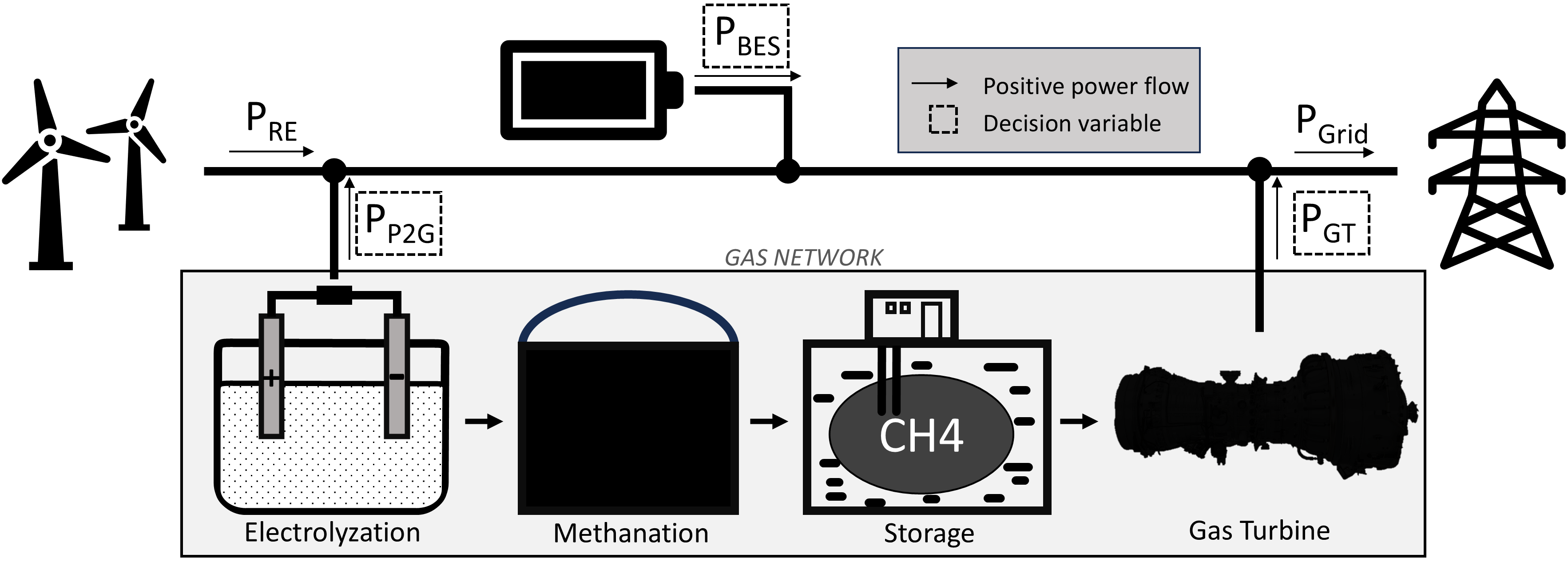}    
		\caption{Simplified plant layout.} 
		\label{fig:layout}
\end{figure*}

\section{Problem Formulation} \label{sec:2_problem_form}

\subsection{System Description}

Figure \ref{fig:layout} provides an overview of the plant modeled in this study. The objective of the optimal dispatch task is to maximize the total profits over the optimization period $T$:
\begin{equation} \label{eq1}
	\max R_{Total} = \sum_{t=0}^{T} R_{Grid, t} - C_{BES, t} - C_{GT, t} - C_{P2G, t}
\end{equation}
where $t$ denotes time steps, $R_{Grid}$ are the revenues from selling power to the utility grid, and $C_{BES}$, $C_{GT}$, $C_{P2G}$ are the cost associated with operating the BES, GT, and P2G system, respectively.

The same notation is used for the power balance constraint, with the addition of the power from renewable energy, $P_{RE}$:
\begin{equation} \label{eq2}
	P_{Grid, t} = P_{RE, t} + P_{BES, t} + P_{GT, t} + P_{P2G, t}
\end{equation}
If the BES is charged, $P_{BES}$ is negative and thus reducing the power sold to the grid. Similarly, $P_{P2G}$ is negative if the P2G system is operating. $P_{P2G}$, $P_{BES}$, and $P_{GT}$ are the decision variables of the dispatch problem.

The revenues from power sales are computed as:
\begin{equation} \label{eq3}
	R_{Grid, t} = P_{Grid, t} \times c_{w, t} \times \Delta t
\end{equation}
where $c_w$ is the wholesale price of electricity. In our case studies, we assume that the modeled plant is a price-taker, able to sell electricity at market prices without influencing them. Constraints \ref{eq4} and \ref{eq5} ensure that grid interaction is limited to selling power and that the BES and P2G system can only be operated on the available renewable power.

\begin{equation} \label{eq4}
	P_{Grid,t} \geq 0
\end{equation}
\begin{equation} \label{eq5}
	P_{BES,t} + P_{P2G,t} \geq - P_{RE, t}
\end{equation}

\subsubsection{Battery Energy Storage Model}
We model the BES operation based on changes in the state of charge $SOC_{BES}$ as the result of charging or discharging activity:
\begin{equation} \label{eq6}
	SOC_{BES, t} = SOC_{BES, t-1} - \eta \frac{P_{BES, t}\times \Delta t}{Cap_{BES}^{max}}
\end{equation}
where $Cap_{BES}^{max}$ is the maximum capacity of the BES and the efficiency $\eta$ is further specified as:

\begin{equation} \label{eq7}
	\eta = 
	\begin{cases}
		1 / \eta_{Dis} &, P_{BES, t} > 0\\
		\eta_{Ch} &, P_{BES, t} \leq 0
	\end{cases}
\end{equation}
depending on charging or discharging.
Our code implementation slightly differs from the simplified representation in equation \ref{eq6} and \ref{eq7} for the case of discharging. We first reduce the SOC according to the discharge action and then apply the discharge efficiency to the difference to calculate the effective power flow. This way, the constraints on maximum discharge ($P^{max}_{BES}$) and minimum SOC ($SOC_{BES}^{min}$) are easier to enforce:
\begin{equation} \label{eq8}
	P^{min}_{BES} \leq P_{BES,t} \leq P^{max}_{BES}
\end{equation}
\begin{equation} \label{eq9}
	SOC_{BES}^{min} \leq SOC_{BES, t} \leq SOC_{BES}^{max}
\end{equation}

We approximate the cost of BES operation by modeling cyclic aging based on the depth of discharge (DOD = 1-SOC):
\begin{equation} \label{eq10}
	C_{BES, t} = \frac{|(1 - SOC_{BES,t})^{k_p} - (1 - SOC_{BES,t-1})^{k_p}|}{2 \times N^{fail}_{100}} \times C_{inv}
\end{equation}
where $k_p$ is the Peukert constant, $N^{fail}_{100}$ is the estimated number of full BES cycles until failure, and $C_{inv}$ are the initial investment cost for purchasing the BES \cite{a227_wang, a228_he}.

\subsubsection{Gas Turbine Model}
The GT is modeled based on a function $F$ that maps power to fuel consumption rates, $q_{Fuel}$:

\begin{equation} \label{eq11}
	q_{Fuel, t} = F(P_{GT, t})
\end{equation}
We will introduce our choice for $F$ along with the case studies in section \ref{sec:4_case_study}. With the fuel being exclusively produced by the P2G system, the only cost we model for GT operation are variable maintenance cost based on a dynamic cost allocation scheme that we have introduced in Ref. \cite{sage2023gt}:
\begin{equation} \label{eq12}
	C_{GT, t} = b_{cycle, t} \times k_{cycle} + b_{oper, t} \times k_{oper}
\end{equation}
where $k_{cycle}$ is a cost factor assigned at GT start-up and computed by dividing the expected lifetime maintenance cost by the expected GT life in cycles: $k_{cycle}=C_{O\&M, L}/L_c$. $k_{oper}$ is a cost factors for continuous operation and computed with the expected GT lifetime in hours: $k_{oper}=C_{O\&M, L}/L_h$. $b_{cycle}$ and $b_{oper}$ are binary indicators. $b_{cycle}$ is 1 whenever the GT starts and 0 otherwise. $b_{oper}$ is calculated as:
\begin{equation} \label{eq13}
	\begin{cases}
		b_{oper} = 1 &, n_h > L_h/L_c\\
		b_{oper} = 0 &, \text{otherwise}
	\end{cases}
\end{equation}
where $n_h$ counts the number of hours the GT operated since it started. As $L_h >> L_c$, this maintenance cost scheme allocates relatively high cost whenever the GT is started. Maintenance cost based on operation time, which are the default in literature, are only assigned for periods of longer, continuous operation.
We have shown in Ref. \cite{sage2023gt, sage2025_b_and_g_preprint} that this effectively discourages frequent GT ramp-ups and prevents unrealistic policies observed when using purely time-based cost allocation schemes. Lastly, GT power is constrained by the maximum possible power of the modeled turbine type:
\begin{equation} \label{eq14}
	0 \leq P_{GT,t} \leq P^{max}_{GT}
\end{equation} 

\subsubsection{Power to Gas System Model}
For the simplified P2G model, we assume that electric power is first used to generate hydrogen in an electrolyzer. Then, the hydrogen and separately bought carbon dioxide are converted to SNG via methanation and stored in a gas storage.
Similar to the BES, the P2G system is modeled based on changes in its SOC due to SNG being added to the gas storage ($q_{genFuel}$) or removed for usage in the GT:
\begin{equation} \label{eq15}
	SOC_{P2G, t} = SOC_{P2G, t-1} + \frac{q_{genFuel, t} - q_{Fuel, t}}{Cap_{P2G}^{max}}
\end{equation}

The quantity of SNG being generated is calculated as:
\begin{equation} \label{eq16}
	q_{genFuel, t} =  - P_{P2G, t} \times \eta_{P2G} \times \alpha_{P2G} \times \Delta t
\end{equation}
where $\eta_{P2G}$ is the combined efficiency of the entire system including the storage and $\alpha_{P2G}$ is a conversion factor from energy to fuel mass. The P2G system's SOC is bounded as $0 \leq SOC_{P2G, t} \leq 1$.
The cost of operating the P2G system consists of two components:
\begin{equation} \label{eq17}
	C_{P2G, t} = 
	\begin{cases}
		0 &, P_{P2G, t} = 0\\
		k_{P2G, fix} + q_{genFuel, t} \times k_{P2G, var} &, P_{P2G, t} < 0
	\end{cases}
\end{equation}
a fixed cost term, $k_{P2G, fix}$, for each time step of operation that represents the degradation cost, and a variable cost term, $ k_{P2G, var}$, based on the produced SNG that represents the cost of purchasing carbon dioxide for methanation.

Finally, if on, the P2G system must operate between a minimum and maximum power level:
\begin{equation} \label{eq18}
	P_{P2G, t} =
	\begin{cases}
		0, & \text{if the P2G system is off} \\
		[P_{P2G}^{min}, P_{P2G}^{max}], & \text{if the P2G system is on}
	\end{cases}
\end{equation}

\subsection{Markov Decision Process}
The Markov Decision Process (MDP) is a popular framework for sequential decision making problems that we utilize in this study. An MDP is characterized by the tuple $(\mathcal{S}, \mathcal{A}, \mathcal{P}, \mathcal{R}, \gamma)$ where $\mathcal{S}$ is the state space, $\mathcal{A}$ is the action space, $\mathcal{P}$ is the transition function that quantifies the probabilities of state transitions, $\mathcal{R}$ is the reward function, and $\gamma$ is the discount factor \cite{a66_sutton2018reinforcement}.

\subsubsection{Reward Function}
For each transition from a state $s_t$ to a new state $s_{t+1}$ as determined by the transition function $\mathcal{P}: \mathcal{S} \times \mathcal{A} \times \mathcal{S} \rightarrow [0,1]$, the agent receives a reward $\mathcal{R}: \mathcal{S} \times \mathcal{A} \rightarrow \mathbb{R}$ for taking action $a_t$ in state $s_t$. The immediate reward at every time step in our environment is $R_{t} = R_{Grid, t} - C_{BES, t} - C_{GT, t} - C_{P2G, t}$. Goal of the agent is to maximize the sum of discounted rewards, which is also called the return:
\begin{equation} \label{eq19}
	G_t = R_{t+1} + \gamma R_{t+2} + \gamma^2 R_{t+3} + ... = \sum_{k=t+1}^{T}\gamma^{k-t-1} R_{k}
\end{equation}
where $\gamma$ is the discount factor and $\gamma \in [0,1]$ \cite{a66_sutton2018reinforcement}. $\gamma$ controls how much the current state-action pair is credited for rewards observed at later time steps. Therefore, with both BES and P2G storing energy for later profits, $\gamma$ is an important hyperparameter in our experiments that we tune carefully.

\subsubsection{State Space}
The state space provides information about the current state of the environment to the agent. In our experiments, the state space is defined as:
\begin{equation} \label{eq20}
	s_t = (P_{RE,t}, c_{w,t}, SOC_{BES, t}, SOC_{P2G, t}, S_{GT,t}, \mathcal{T}_t)
\end{equation}
and includes the available renewable power, the wholesale price of electricity, and the SOCs of BES and P2G system. $S_{GT}$ indicates the operating state of the GT. $S_{GT}$ is 0 if the GT is off, 1 if it started recently (before reaching the $L_h/L_c$ ratio), and 2 if it started longer ago (after passing the $L_h/L_c$ ratio). $\mathcal{T}$ is a placeholder for different temporal features. In our previous work, we have shown that sine- and cosine-encoded time counters, for example for the hour of the day, lead to significant performance improvements \cite{sage2025_bes_comprehensive}. The composition of $\mathcal{T}$ varies between case studies and will be explained in section \ref{sec:4_case_study}.

\subsubsection{Action Space}
The action space is three-dimensional, with one dimension each for GT, P2G system, and BES. All dimension are continuous, with ranges of [0, $P_{GT}^{max}$] for the GT, [$P_{P2G}^{max}$, 0] for the P2G system, and [$P_{BES}^{min}$, $P_{BES}^{max}$] for the BES. We scale all dimensions to a [-1,1] range that we keep fixed during DRL training. To prevent constraint violating actions, such as charging an already full BES, we make use of a safety mechanism that corrects the actions chosen by the DRL agent to the nearest possible action.

Some DRL algorithms either require discrete action spaces or can operate on both discrete and continuous action spaces. In preliminary experiments, we have tested several DRL algorithms and action space variations and observed better performance with discrete action spaces. In the introduced environment, fine-grained control is not required due to the absence of demands. Instead, the type of decision, such as starting the GT or not, is more important. A coarse discretization can help to simplify the environment, while the safety mechanism corrects for imprecision. For optimal performance, we tune the level of discretization for each dimension separately and report our findings in the results.


\section{Methodology} \label{sec:3_method}

\subsection{Deep Reinforcement Learning}

We apply two DRL algorithms to the hybrid energy system introduced in the previous section: DQN and PPO. Our choice is motivated by the reviewed literature and our own experience in earlier work \cite{sage2023gt, sage2025_bes_comprehensive, sage2024_DRL_forecasting, sage2025_b_and_g_preprint}. In the following, we will briefly introduce the core concepts of both algorithms. For further details and pseudocode, the reader is referred to the original papers \cite{a79_mnih2015human, a121_schulmanPPO}.

DQN \cite{a79_mnih2015human} is a value-based method that trains a neural network, called Q-network, to learn the state-action (or Q-) function using the equations:
\begin{equation} \label{eq23}
	\delta_i = r + \gamma \max_{a'} Q(s', a'; \theta^-)
\end{equation}
\begin{equation} \label{eq24}
	Q(s,a; \theta) = Q(s,a; \theta) + \alpha [\delta_i - Q(s,a; \theta)] 
\end{equation}
where $r$ is the reward, $\theta$ are the parameters of the Q-network, $\alpha$ is the learning rate, and $s'$, $a'$ denote the next state and action, respectively. $\theta^-$ are the parameters of the target network, a copy of the Q-network that is updated periodically with the parameters $\theta$ to improve stability during training.

During training, the $\epsilon$-greedy action selection strategy balances the trade-off between exploration and exploitation. With $\epsilon$ probability, a random action is chosen, and with 1-$\epsilon$ probability, the action that maximizes the next Q-value is chosen. We anneal $\epsilon$ during training to favor exploration initially and exploitation later. After training, the optimal policy $\pi^*$ can be obtained by selecting those actions that maximize the Q-values for each state:
\begin{equation} \label{eq_m2}
	\pi^*(s) = \arg \max_a Q(s,a, \theta).
\end{equation}
DQN is an off-policy method that utilizes a replay buffer to store experiences. These experiences are randomly sampled from the buffer and replayed in order to compute gradient updates. This approach increases the sample efficiency and stability of updates.

PPO \cite{a121_schulmanPPO} is an actor-critic method, incorporating an actor, which proposes actions given states, and a critic, which evaluates the proposed actions by estimating the value function. In the discrete version that we employ in our study, the actor learns the parameters of a categorical distribution - one output for each possible action. During training, actions are sampled from this distribution to ensure exploration. Conversely, during evaluation, the action with the highest probability is selected to maximize performance.

A defining feature of PPO is its clipped surrogate objective function:
\begin{equation} \label{eq_m3}
	L^{CLIP}(\theta) = \mathbb{E}_t \left[\min\left(r_t(\theta) \hat{A}_t, \text{clip}(r_t(\theta), 1 - \epsilon, 1 + \epsilon) \hat{A}_t\right)\right]
\end{equation}
where $\theta$ denotes the parameters of the policy, $r_t(\theta)$ represents the probability ratio between the current and old policy for the action taken, and $\hat{A}_t$ is the advantage at time $t$, indicating how much better an action is compared to the average action at a given state. The clipping mechanism, controlled by $\epsilon$, mitigates the risk of excessively large policy updates and thereby enhances stability.

As on-policy method, in PPO the experiences used for training the policy are generated by the same policy currently under evaluation. Thus, PPO is always updated with the most recent data and does not reuse past experiences.

\subsection{Methods for Improved Power-to-Gas Control} \label{sec:modifications}
Economically dispatching energy storages such as BES and P2G systems with GTs poses a challenge for DRL methods. The agent must learn to charge the storage - causing operational cost and potentially missing out on immediate sales - in order to achieve higher rewards later. Especially for P2G systems, with high cost, low efficiencies, and long periods between charge and discharge, learning to connect actions and rewards is challenging.
Throughout all the conducted case studies, we observed DRL struggling with P2G deployment, often leading to policies in which both P2G system and GT remained unused. We have therefore derived different strategies to cope with this issue, which we will introduce in the following.

\subsubsection{Forecasting Future Information}
A straightforward strategy to improve DRL performance is to provide additional information to the agent, particularly regarding the future of uncertain variables. In our environment, the wholesale price of electricity is the main factor for decision-making. Therefore, we experiment with adding price forecasts to the state space. For simplicity and to reduce computational cost, we utilize perfect forecasts (ground truth) instead of predictions. We have thoroughly investigated the effect of price forecasts with deep learning models for energy arbitrage with BES and DRL in Ref. \cite{sage2024_DRL_forecasting}. Our study showed significant performance improvements where forecasts of different time horizons, ranging from 1 hour to 24 hours, were grouped. Furthermore, we demonstrated that the grouping of different predictions, despite large forecasting errors, leads to comparable results as when perfect forecasts are used. Building upon this prior work, we utilize ground truth price forecasts for 1, 2, 3, 6, 12, 18, and 24 hours to investigate if similar performance gains are feasible with DRL-controlled P2G systems.

\subsubsection{Reward Penalty: Low P2G SOC}
A different approach is to penalize the agent for certain behavior by subtracting extra terms from the reward function. We experiment with two such penalties with the motivation to encourage P2G usage. The first penalty is assigned for low SOCs of the P2G system, and it depends on how much SNG is available. To quantify this penalty, we first define an upper SOC penalty limit, $SOC^{p,max}_{P2G}$, beyond which no penalty is incurred. Then, the penalty is calculated as:
\begin{equation} \label{eq_m4}
	p_t = \omega \times \max(\frac{SOC^{p,max}_{P2G} - SOC_t}{SOC^{p,max}_{P2G}}, 0)
\end{equation}
where $\omega$ is the penalty weight. Both $SOC^{p,max}_{P2G}$ and $\omega$ are parameters that we tune along with the model specific hyperparameters to identify the best values. The idea behind this penalty design is that the agent is incentivized to maintain an SOC above the defined threshold, thereby ensuring that the GT remains able to capitalize on electricity price spikes at any time. 

\subsubsection{Reward Penalty: P2G Inactivity}
For the second penalty, we directly aim at increasing P2G system activity during periods of low electricity prices. The mechanism of this penalty is shown in pseudocode in Algorithm \ref{algo1}. The algorithm stores a running mean of electricity prices that it compares to the current price after applying a threshold. If the current price falls below the threshold price and the P2G system remains unused despite the availability of sufficient renewable energy, then the full penalty $\omega$ is assigned.

This penalty function adds three parameters ($\omega$, $\tau$, and $\alpha$) that we tune along with the model-specific hyperparameters. With high cost and low round-trip efficiencies, P2G system operation can only be economical if SNG is generated at much lower prices than burnt. With the inactivity penalty and its parameters we intend to facilitate the corresponding learning process.

\begin{algorithm}
	\caption{P2G inactivity penalty based on electricity price}
	\label{algo1}
	\begin{algorithmic}[1]
		\State \textbf{Initialize:} penalty weight $\omega$, threshold $\tau$, learning rate $\alpha$, mean wholesale electricity price $\overline{c_w}$	
		\For{each step $t$ of one episode}:
		\State $\overline{c_w} = \overline{c_w} + \alpha \times (c_{w,t} - \overline{c_w})$
		\Comment{Update running mean}
		\State $c_{w, \tau} = \overline{c_w} \times \tau$
		\Comment{Get threshold price}
		\If{$c_{w,t} <= c_{w, \tau} \text{ and } P_{P2G, t} == 0 \text{ and } P_{RE, t} >= -P^{min}_{P2G}$}
		\State $r = r - \omega$
		\Comment{Update reward}
		\EndIf
		\EndFor
	\end{algorithmic}
\end{algorithm}

\subsubsection{Reward Shaping: P2G Cost Attribution}

\begin{figure}[h]
		\centering
		\includegraphics[width=\linewidth]{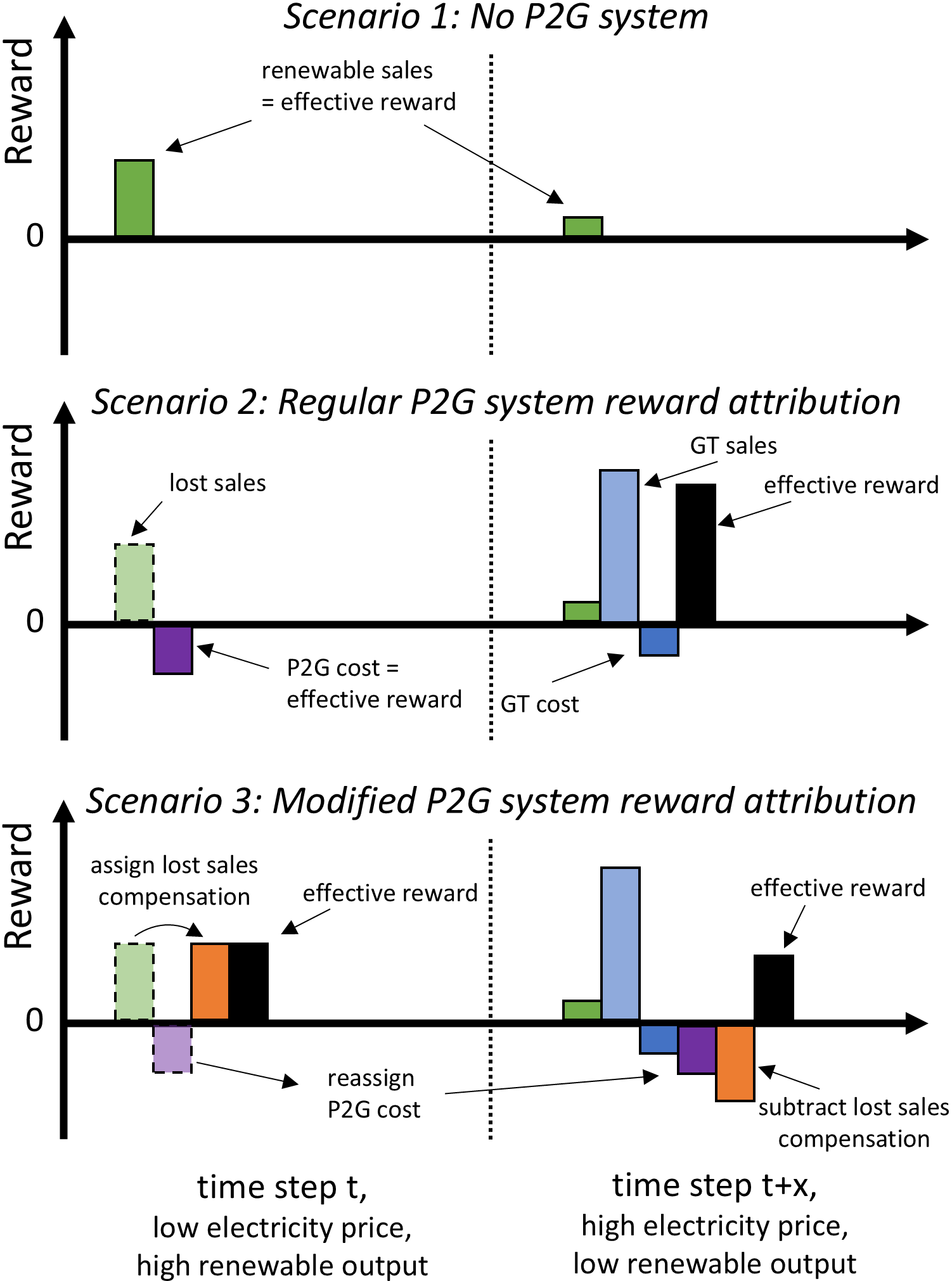}    
		\caption{A simplified visualization of the proposed reward attribution for P2G system cost and lost sales for two time steps.}
		\label{fig:reward_attribution}
\end{figure}

A different approach that also affects the reward function is a changed attribution of the opportunistic cost of P2G usage, namely the missed out sales and the operational cost of the P2G system. Figure \ref{fig:reward_attribution} provides a visual example of how we re-attribute rewards, leaving the BES out for better clarity. If no P2G system usage occurs, the renewable energy sales correspond to the effective rewards. In the regular setup introduced in section \ref{sec:2_problem_form}, charging the P2G system is seemingly unattractive due to lost sales of renewables and P2G system cost. Instead, rewards are significantly higher if the generated SNG is burnt in the GT during high-price periods. The modified reward attribution consists of two components: 1) P2G system cost are saved internally during operation and reassigned to those time steps when fuel is released from storage; 2) For lost sales of renewables, a compensation term is introduced that leaves the original reward unchanged. This compensation term is subtracted from rewards later on when fuel is being discharged. Scenario 3 in Figure \ref{fig:reward_attribution} shows this approach for both terms. The resulting rewards are higher during P2G system operation and lower when fuel is burnt in the GT. A different interpretation of this method could be that the agent pays for fuel usage instead of fuel generation, as it would when connected to the gas grid.

Algorithm \ref{algo2} puts the proposed reward attribution in pseudocode. An important implementation detail is the usage of the SOC deduction ratio $\rho$ which allows to assign lost sales compensations and P2G system cost proportional to the fuel usage, enabling assignments across multiple time steps. The sum of rewards over an entire episode remains identical as in the case of regular cost attribution if all the fuel is burnt by the episode end.

\begin{algorithm}
	\caption{Modified cost attribution for P2G system}
	\label{algo2}
	\begin{algorithmic}[1]
		\State \textbf{Initialize:} sum of P2G cost: $\sum_{Cost}=0$, sum of lost sales = $\sum_{LS}=0$
		\For{each step $t$ of one episode}:
		\If{$P_{P2G, t} < 0$}:
			\Comment{P2G system is operating.}
			\State{$\sum_{Cost}= \sum_{Cost} + C_{P2G, t}$}
			\State{$C_{P2G, t}=0$}
			\Comment{Overwrite internal $C_{P2G, t}$}
			\State{$C_{Comp, t} = -P_{P2G,t} \times c_{w,t}$}
			\Comment{Sales compensation}
			\State{$\sum_{LS}= \sum_{LS} + C_{Comp, t}$}
			\State{$r = r + C_{Comp, t}$}
			\Comment{Modify reward}
			\EndIf
		\If{$q_{Fuel, t} > 0$}:
			\Comment{GT is operating.}
			\State{$\rho = (SOC_{P2G, t-1} - SOC_{P2G, t}) / SOC_{P2G, t-1}$}
			\State{$C_{P2G, t} = \sum_{Cost} \times \rho$}
			\Comment{Overwrite internal $C_{P2G, t}$}
			\State{$\sum_{Cost} = \sum_{Cost} - C_{P2G, t}$}
			\State{$C_{Comp, t} = \sum_{LS} \times \rho$}
			\State{$\sum_{LS} = \sum_{LS} - C_{Comp, t}$}
			\State{$r = r - C_{Comp, t} - C_{P2G, t}$}
			\Comment{Modify reward}
			\EndIf
		\EndFor
	\end{algorithmic}
\end{algorithm}


\section{Case Study} \label{sec:4_case_study}

We apply the proposed methodology to three case studies with increasing complexity. All three case studies control the same hybrid energy system in an hourly resolution, but with different episode lengths and on different datasets. Case studies 1 (CS1) and 2 (CS2) are based on synthetic renewable energy and electricity price data and model episodes of 24 hours and 168 hours, respectively. At the end of the episodes, significant spikes in electricity prices occur, which can be better capitalized on if the P2G system was deployed in previous time steps. Thus, with these two case studies we benchmark the DRL agents’ ability to dispatch P2G systems if the corresponding rewards occur up to one day and up to one week later.

\begin{figure}
	\centering
		\includegraphics[width=\linewidth]{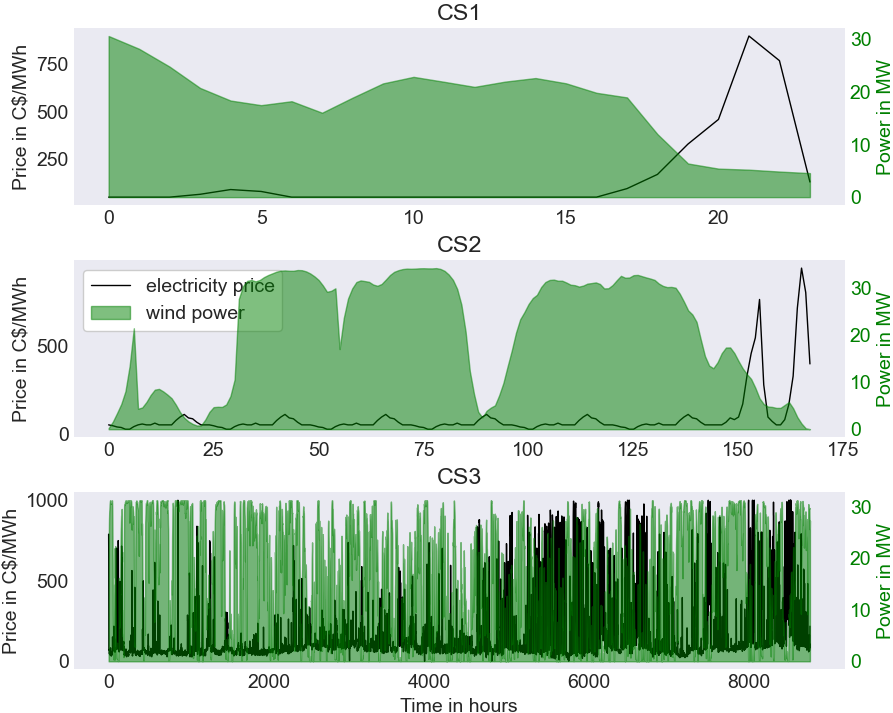}    
		\caption{Time-series data for wind power generation and wholesale electricity prices used in case studies CS1, CS2, and CS3.}
		\label{fig:cs}
\end{figure}

The third case study (CS3) spans over the year of 2022 (8760 time steps) and simulates plant operation in Southern Alberta, Canada at 49.05\degree N, 112.75\degree W. We obtained wholesale electricity prices from the Alberta Electric System Operator (AESO) \cite{aeso} and renewable energy data from renewables.ninja \cite{a206_STAFFELL2016_wind} modeling the operation of seven 4.5 MW wind turbines (Siemens Gamesa SG 4.5 145) at 107.5m hub height.
The time-series data for the three case studies is depicted in Figure \ref{fig:cs}. The data generated for CS1 and CS2 matches the magnitudes of the data obtained for CS3 so that the same plant configuration can be applied. CS3 shows frequent but irregular price spikes with different magnitudes. While the mean price is 163 C\$/MWh, the median is 84 C\$/MWh and the standard deviation is 199 C\$/MWh. For CS1 and CS2, we do not use additional time features (see $\mathcal{T}$ in Eq. \ref{eq20}). For CS3, we utilize sine- and cosine-encodings for the hour of the day, week of the year, and month of the year.

To model the fuel consumption of the GT, we utilize the following piecewise linear equation:
\begin{equation} \label{eq_cs1}
	\begin{cases}
		q_{Fuel, t} = 0 &, P_{GT, t} = 0\\
		q_{Fuel, t} = 700P_{GT, t} + 1550 &, 0 < P_{GT, t} \leq 1 MW\\
		q_{Fuel, t} = 360P_{GT, t} + 2200 &, P_{GT, t} \geq 1 MW\\
	\end{cases}
\end{equation}
where $q_{Fuel}$ is measured in pounds per hour and $P_{GT}$ in MW. We have used the same approximation in our previous study \cite{sage2025_b_and_g_preprint}, along with the other parameters listed for the GT in Table \ref{tbl:params}. While we do not incorporate ramping constraints due to the coarse temporal resolution of our case studies, we correct fuel consumption and power output for hours in which the GT is started. For this correction, we assume 20 minutes of start-up time until the desired power level is reached and an average fuel flow of 1,200 pounds per hour during start-up. As a consequence, to fully exploit price spikes, the agent must start the GT before the highest price is reached.

The parameters used on all three case studies for BES and P2G system are listed in Table \ref{tbl:params}. The BES can be fully charged and discharged in approximately two hours. This limits its ability to capture renewable energy during low price periods. The P2G system storage has a significantly larger capacity and could fuel the GT for 72 hours at maximum power if fully charged. This large capacity is possible as we assume the availability of a depleted gas well for the storing of the SNG. We further assume efficiencies of 75\% for the electrolyzer \cite{w28_SEelyzer}, 83\% for methanation \cite{a256_GHAIB2018433} and 90\% for compression and storage losses, leading to a combined efficiency of 56\%. We compute $k_{P2G, fix}$ based on investment cost of one million C\$ per installed MW \cite{a257_ALBREIKI} and a equipment lifetime of 100,000 hours \cite{w28_SEelyzer}. For $k_{P2G, var}$, we assume a purchasing price of C\$ 12.50 per tonne of CO2 under a long-term contract with an ammonia producer in Alberta \cite{w27_co2cost}, and a consumption of 2.7 kg CO2 per kg of CH4. Finally, $\alpha_{P2G}$ is computed using the lower heating value for CH4 of 50 MJ/kg.

With these parameters we can do a simple comparison for the break even price of BES and P2G-GT. If charged at the median price of 84 C\$/MWh for one hour, the BES must discharge its energy at a price of 158 C\$/MWh to break even. The P2G on the other hand, operates for approximately five hours at maximum capacity to produce the SNG needed to operate the GT close to its maximum capacity for one hour. Taking the occurring cost for P2G and GT into consideration, the break even price is as high as 504 C\$/MWh. Along with the delayed rewards, this example highlights the difficult economics of P2G system operation.

The codebase is implemented in Python 3.11, using the libraries Gymnasium \cite{gymnasium} for the environment, stable-baselines3 \cite{stable-baslines3} for the DRL models, and Optuna \cite{optuna} for hyperparameter tuning. Due to the known issues with DRL stability, we dedicate significant resources to hyperparameter optimization. We first tune both DRL models without the introduced methods for improved P2G control and then conduct another round of tuning including these methods and their respective parameters for around 1,000 trials on each case study. Then, based on the best parameters found, we conduct a sensitivity analysis for the proposed methods. For a fair comparison, all methods that manipulate the reward function are only applied during training and deactivated for evaluation.

As non-DRL benchmarks, we utilize the cross-entropy method (CEM) and mixed-integer quadratic programming (MIQP). The CEM is a heuristic approach that seeks to find the optimal parameters of a policy network in an evolutionary approach \cite{cem}. While the CEM serves as lower bound for the DRL models, the MIQP method, implemented using Gurobi \cite{gurobi}, determines the optimal solution given full access to the environment and future states.

\begin{table}[]  
	\centering
		\caption{Parameters of the BES, GT, and P2G system.}  
		\label{tbl:params}  
		\begin{tabular}{l|l}  
			\hline
			\textbf{Parameter}  & \textbf{Value}  \\ 
			\hline  
			\multicolumn{2}{c}{\textit{Battery Energy Storage}} \\ 
			\hline  
			$Cap^{max}_{BES}$ & 50 MWh \\  
			$SOC^{min}_{BES}$/ $SOC^{max}_{BES}$ & 0.1 / 0.9  \\  
			$P^{min}_{BES}$ / $P^{max}_{BES}$ & -20 / 20 MW \\  
			$\eta_{Ch}$/$\eta_{Dis}$ & 0.92\\  
			Peukert constant, $k_p$ & 1.14 \cite{a229_tran_bes}\\  
			Cycles to failure, $N^{fail}_{100}$ & 6,000 \cite{a230_cheng_bes} \\  
			Investment cost, $C_{inv}$ & 300,000 C\$/MWh \cite{w23_cole_cost}\\  
			\hline  
			\multicolumn{2}{c}{\textit{Gas Turbine}} \\  
			\hline  
			$P^{max}_{GT}$ & 32.6 MW\\
			GT life in cycles, {$L_c$} & 26,000 \cite{w18_turbomachinery}\\  
			GT life in hours, {$L_h$} & 200,000 \cite{w18_turbomachinery}\\  
			lifetime O\&M cost, {$C_{O\&M,L}$} & C\$33.000.000 \cite{w17_energy2020capital}\\  
			\hline  
			\multicolumn{2}{c}{\textit{Power-to-Gas System}} \\  
			\hline
			$Cap^{max}_{P2G}$ & 1 M lbs \\
			$P^{min}_{P2G}$ / $P^{max}_{P2G}$ & -12 / -30 MW \\
			$\eta_{P2G}$ & 0.56 \cite{w28_SEelyzer, a256_GHAIB2018433}\\  
			$\alpha_{P2G}$ & 158.73 lbs/MWh \\
			$k_{P2G, fix}$ & 300 C\$/h \cite{w28_SEelyzer, a257_ALBREIKI}\\
			$k_{P2G, var}$ & 0.03375 C\$/kg \cite{w27_co2cost} \\
			\hline
		\end{tabular}  
\end{table}  



\section{Results and Discussion} \label{sec:5_results}

\begin{figure}
	\centering
		\includegraphics[width=\linewidth]{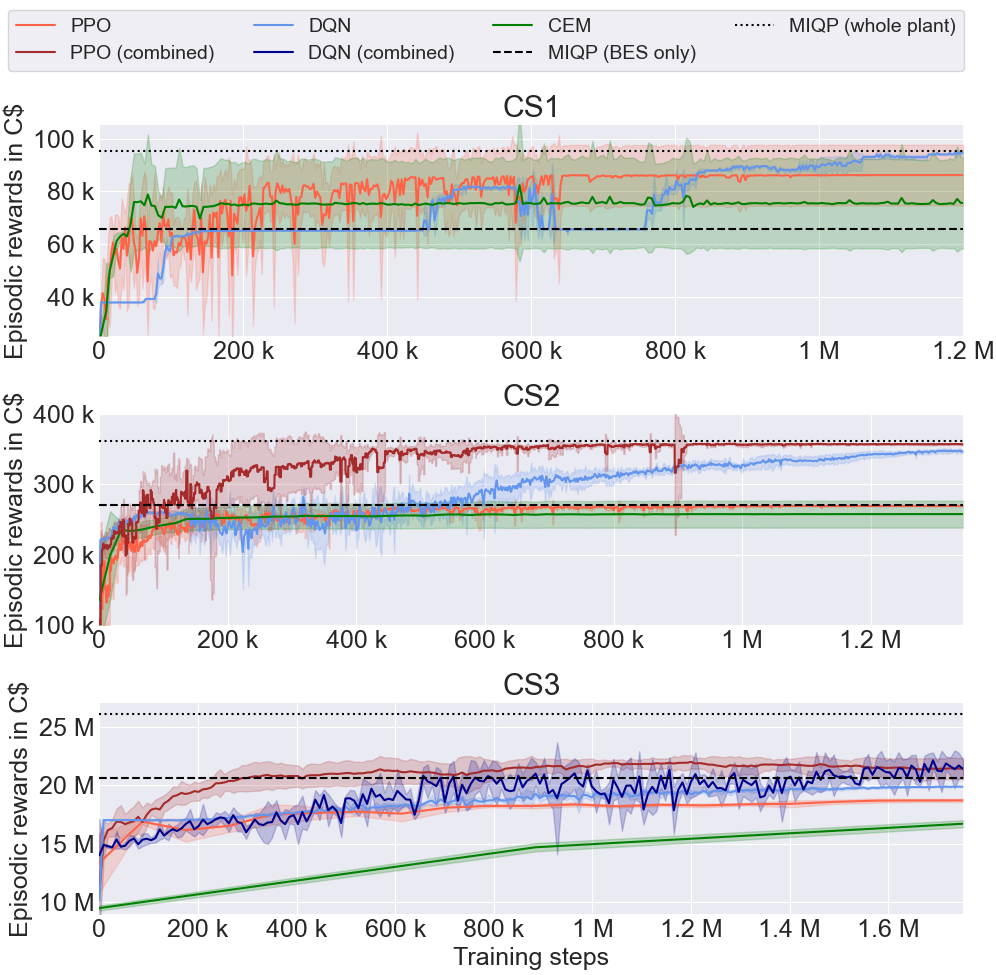}    
		\caption{Training progress of tested models measured in accumulated episodic rewards over training steps. The shaded areas show $\pm$ one standard deviation for five runs with different random seeds.}
		\label{fig:training}
\end{figure}

\begin{table*}[h!]  
		\centering
		\caption{Summary of results for all three case studies. Episodic rewards and standard deviations are rounded to the nearest thousand. Regarding variants, 'Base' refers to the DRL algorithm without the modifications proposed in section \ref{sec:modifications}, 'Forecast' to experiments with electricity price forecasts, 'Cost-Attr.' to the delayed attribution of P2G system cost, 'SOC-P' to penalties for low P2G-SOCs, 'INA-P' to penalties for P2G inactivity, and 'Combined' for the best combination found out of all modifications.}  
		\label{tbl:results}
		\small 
		\renewcommand{\arraystretch}{1} 
		\begin{tabular}{|l|l|l|r|c|c|c|c|c|} 
			\toprule  
			\textbf{\makecell{Case\\Study}} & \textbf{Algorithm} & \textbf{Variant} & {\textbf{\makecell{Episodic\\Reward (kC\$)}}} & {\textbf{\makecell{Num. GT\\Starts}}} & {\textbf{\makecell{Num. GT\\ Hours}}} & {\textbf{\makecell{Num. P2G\\ Hours}}} & {\textbf{\makecell{Num. BES\\ Charge}}} & {\textbf{\makecell{Num. BES\\ Discharge}}} \\ \midrule  
			CS1 & DQN & Base & 94 $\pm$ 1 & 1 & 3 & 16 & 2 & 2 \\  
			& PPO & Base & 86 $\pm$ 10 & 0.8 & 1.6 & 9.6 & 2 & 2 \\  
			& MIQP &  & 95 $\pm$ 0 & 1 & 3 & 15 & 3 & 2 \\  
			\addlinespace  
			CS2 & DQN & Base & 347 $\pm$ 2 & 2 & 7.6 & 39 & 31 & 15 \\  
			& PPO & Base & 269 $\pm$ 1 & 0 & 0 & 0 & 20 & 9 \\  
			&  & Forecast & 268 $\pm$ 2 & 0 & 0 & 0 & 23 & 11 \\  
			&  & Cost-Attr. & 267 $\pm$ 0 & 0 & 0 & 0 & 14 & 4 \\  
			&  & SOC-P & 349 $\pm$ 16 & 2 & 8 & 39 & 13 & 5 \\  
			&  & INA-P & 339 $\pm$ 5 & 3 & 13 & 73 & 21 & 11 \\  
			&  & Combined & 357 $\pm$ 1 & 2 & 9 & 48 & 17 & 7 \\  
			& MIQP &  & 361 $\pm$ 0 & 2 & 11 & 49 & 30 & 8 \\  
			\addlinespace  
			CS3 & DQN & Base & 19,896 $\pm$ 39 & 6 & 6 & 10 & 1730 & 823 \\  
			&  & Forecast & 20,809 $\pm$ 17 & 37 & 73 & 287 & 1481 & 684 \\  
			&  & Cost-Attr. & 20,247 $\pm$ 386 & 36 & 90 & 488 & 1370 & 711 \\  
			&  & SOC-P & 20,944 $\pm$ 115 & 100 & 144 & 712 & 1442 & 702 \\  
			&  & INA-P & 20,304 $\pm$ 86 & 60 & 89 & 271 & 1398 & 676 \\  
			&  & Combined & 22,292 $\pm$ 326 & 133 & 428 & 2520 & 1355 & 628 \\  
			\addlinespace  
			& PPO & Base & 18,716 $\pm$ 107 & 16 & 16 & 25 & 1748 & 864 \\  
			&  & Forecast & 19,615 $\pm$ 112 & 23 & 25 & 44 & 1272 & 568 \\  
			&  & Cost-Attr. & 18,554 $\pm$ 1029 & 37 & 104 & 575 & 484 & 239 \\  
			&  & SOC-P & 20,891 $\pm$ 2081 & 106 & 268 & 1473 & 948 & 372 \\  
			&  & INA-P & 19,493 $\pm$ 634 & 171 & 372 & 1786 & 946 & 425 \\  
			&  & Combined & 22,411 $\pm$ 377 & 95 & 397 & 2517 & 1101 & 454 \\  
			\addlinespace  
			& MIQP &  & 26,087 $\pm$ 0 & 106 & 549 & 2668 & 2610 & 747 \\  
			\bottomrule  
		\end{tabular}  
\end{table*}  

\begin{figure*}[h!]
		\centering
		\includegraphics[width=\textwidth]{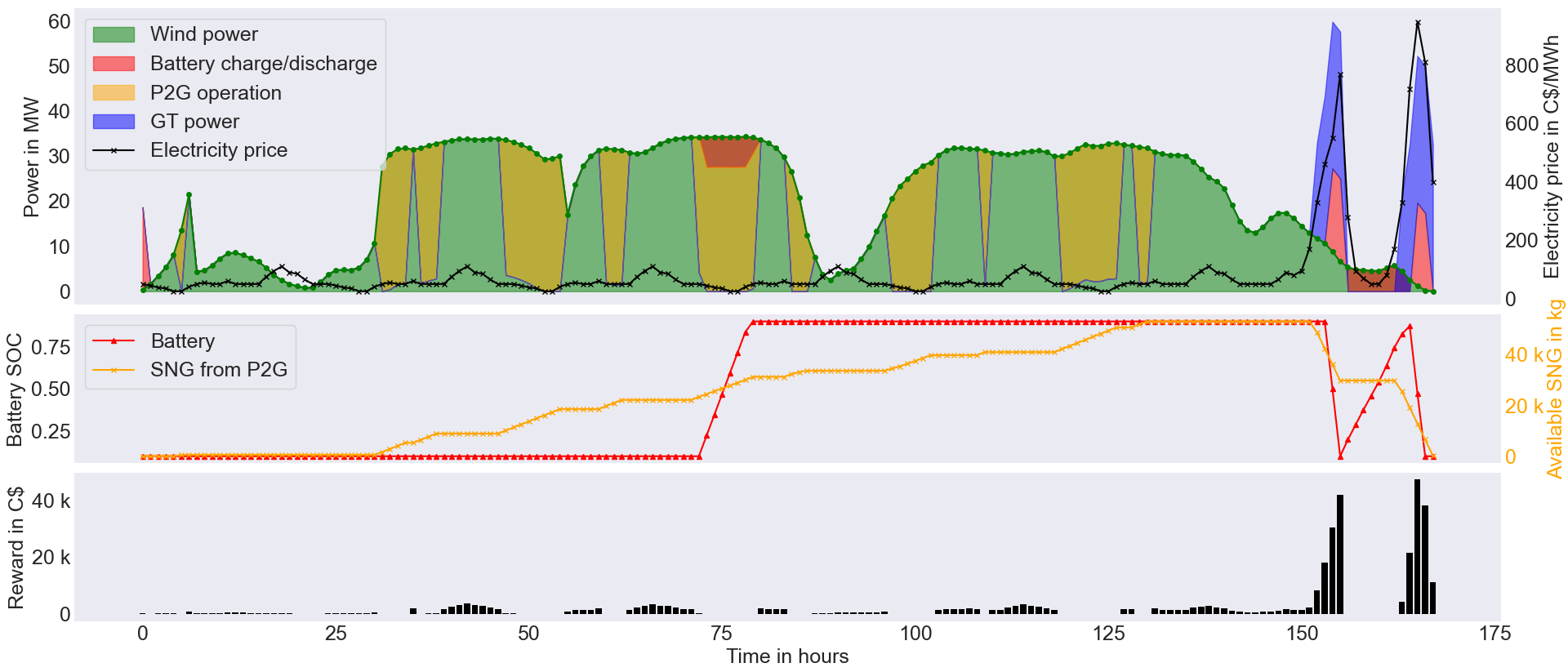}    
		\caption{Best policy learned by PPO on CS2, using the combination of proposed modifications. (TOP) The dispatch decisions with P2G system, BES, and GT at every time step of the simulated week, alongside the wholesale electricity prices. (MIDDLE) Changes in BES SOC and available SNG from P2G operation. (BOTTOM) The immediate reward at each time step.} 
		\label{fig:168sample}
\end{figure*}

The key results of our experiments on the three case studies are listed in Table \ref{tbl:results}, comparing different variants of both DRL models and MIQP regarding rewards and learned policies. In Figure \ref{fig:training}, we further show the training progress of selected models and variants. The figure includes MIQP variants that dispatch the BES only to visually identify DRL policies without P2G usage. On CS1, both DQN and PPO obtained high rewards without the usage of the proposed modifications. DQN closely approached the optimal policy, while PPO scored lower as one of the five independent runs failed to surpass the BES-only benchmark. Similarly, CEM struggled to reliably reach high rewards across independent runs.

On CS2, DQN again reached a high-performing policy without additional modifications. PPO, though, failed to make use of P2G system and GT. This however changed dramatically as we introduced the proposed modifications. While using forecasts and attributing P2G cost at fuel usage did not improve performance, the penalties on low SOCs and P2G system inactivity each led to the desired increase in P2G/GT activity and subsequently in rewards. The combination of modifications further improved PPO, leading to a policy that nearly matched the optimum. Figure \ref{fig:168sample} shows the policy learned by this modified PPO agent. High rewards were obtained as the agent learned to convert a significant quantity of the wind energy to SNG, especially when prices were low, and then utilize it to operate the GT during both price peaks. For the BES, the learned dispatch strategy is more flexible and allowed for a recharge between both peaks.

Both DQN and PPO struggled with the significantly more complex CS3, scoring less than the BES-only benchmark and barely using the P2G system. For DQN, each modification measurably increased P2G system usage and rewards, with SOC-based penalties yielding the most improvement. When combined, the proposed modifications further raised rewards to a 12\% increase compared to the base case. For PPO, we observed a similar behavior: the SOC-based penalty was the most influential modification, while best results were obtained for the combination of all modifications (+20\%). 

In all experiments, we obtained best results when parameterizing the SOC penalty as $\omega=1000$ and $SOC^{p,max}_{P2G}=0.01$. This $SOC^{p,max}_{P2G}$ enables the GT to generate around 22 MWh of electricity if burnt in one hour. Thus, the penalty worked best when motivating the agent to store some fuel to be used during a single price spike. For the inactivity-based penalty, the best parameters found were $\omega=1000$, $\alpha=0.02$, $\tau=0.7$. Delaying the attribution of the P2G system’s opportunistic cost barely influenced performance. Our results indicated that a stronger incentive for P2G usage is necessary than assigning cost during fuel consumption. Despite using the ground truth, the DRL variants with access to forecasts only yielded minor improvements. It is likely that for a bigger impact, longer forecasting horizons are required. However, as accurate long-term forecasts are difficult to obtain, this modification might be of limited use for P2G dispatch. Regarding discretization, both DRL models worked best in the coarsest possible set-up, with two action choices each for GT and P2G and three action choices for the BES.


\section{Conclusions} \label{sec:6_conclusions}

In this study, we have explored the applicability of DRL to the long-term dispatch of hybrid energy systems with P2G technology. As the complexity of the case studies increased, we have observed that the challenges associated with deploying DRL also grew. Despite significant efforts in tuning, it is not guaranteed that DRL will learn effective and stable policies. We have proposed modifications to the problem formulation that have proven to be crucial, enabling DRL agents to successfully learn the economic dispatch of P2G system and GT. Among the proposed modifications, penalties for low SOCs of the gas storage and penalties for P2G system underutilization during low-price periods have performed particularly well. 

Interestingly, these findings stand in contrast to our previous work on reward penalties for BES dispatch, where similar penalties barely improved results or even caused undesirable behavior \cite{sage2025_bes_comprehensive}. The key difference between P2G/GT dispatch and BES dispatch is that off-the-shelf DRL agents are capable of learning BES dispatch without such modifications. In fact, in all three case studies, the rewards of pure BES dispatch by DRL nearly matched those of the optimum policies.

Our experiments with combined modifications suggest the presence of positive cross-effects. Further investigation into these effects and additional reward shaping methods represents a promising direction for future research.



\section*{Acknowledgments}
This work is supported by MITACS grant number IT13369 and Siemens Energy. Manuel Sage is recipient of the Fonds de recherche du Québec - Nature et technologies (FRQNT) Doctoral Training Scholarship. The authors thank Digital Research Alliance of Canada for providing computational resources.



\bibliographystyle{asmeconf}  
\bibliography{references}

\begin{thebibliography}{10}
\newcommand{\enquote}[1]{``#1''}
\providecommand{\url}[1]{\texttt{#1}}
\providecommand{\urlprefix}{URL }
\expandafter\ifx\csname urlstyle\endcsname\relax
  \providecommand{\doi}[1]{DOI \discretionary{}{}{}#1}\else
  \providecommand{\doi}{DOI \discretionary{}{}{}\begingroup
  \urlstyle{rm}\Url}\fi
\providecommand{\eprint}[2][]{\urlprefix\url{#1#2}}

\bibitem{sage2024_DRL_forecasting}
Sage, Manuel, Campbell, Joshua and Zhao, Yaoyao~Fiona.
\newblock \enquote{Enhancing Battery Storage Energy Arbitrage With Deep
  Reinforcement Learning and Time-Series Forecasting.}
\newblock \textit{ASME 2024 18th International Conference on Energy
  Sustainability}. 2024.
\newblock \doi{10.1115/ES2024-130538}.

\bibitem{sage2025_bes_comprehensive}
Sage, Manuel and Zhao, Yaoyao~Fiona.
\newblock \enquote{Deep reinforcement learning for economic battery dispatch: A
  comprehensive comparison of algorithms and experiment design choices.}
\newblock \textit{Journal of Energy Storage} Vol. 115 (2025): p. 115428.
\newblock \doi{https://doi.org/10.1016/j.est.2025.115428}.

\bibitem{a225_huang}
Huang, Bin and Wang, Jianhui.
\newblock \enquote{Deep-Reinforcement-Learning-Based Capacity Scheduling for
  PV-Battery Storage System.}
\newblock \textit{IEEE Transactions on Smart Grid} Vol.~12 No.~3 (2021): pp.
  2272--2283.
\newblock \doi{10.1109/TSG.2020.3047890}.

\bibitem{a213_CAO_energyarbitrage}
Cao, Jun, Harrold, Dan, Fan, Zhong, Morstyn, Thomas, Healey, David and Li,
  Kang.
\newblock \enquote{Deep Reinforcement Learning-Based Energy Storage Arbitrage
  With Accurate Lithium-Ion Battery Degradation Model.}
\newblock \textit{IEEE Transactions on Smart Grid} Vol.~11 No.~5 (2020): pp.
  4513--4521.
\newblock \doi{10.1109/TSG.2020.2986333}.

\bibitem{a237_HARROLD2022121958}
Harrold, Daniel~J.B., Cao, Jun and Fan, Zhong.
\newblock \enquote{Data-driven battery operation for energy arbitrage using
  rainbow deep reinforcement learning.}
\newblock \textit{Energy} Vol. 238 (2022): p. 121958.
\newblock \doi{10.1016/j.energy.2021.121958}.

\bibitem{a253_ZHANG2020113063}
Zhang, Bin, Hu, Weihao, Li, Jinghua, Cao, Di, Huang, Rui, Huang, Qi, Chen, Zhe
  and Blaabjerg, Frede.
\newblock \enquote{Dynamic energy conversion and management strategy for an
  integrated electricity and natural gas system with renewable energy: Deep
  reinforcement learning approach.}
\newblock \textit{Energy Conversion and Management} Vol. 220 (2020): p. 113063.
\newblock \doi{https://doi.org/10.1016/j.enconman.2020.113063}.

\bibitem{a179_teng2021}
Teng, Xinyuan, Long, Huan and Yang, Luoxiao.
\newblock \enquote{Integrated Electricity-Gas System Optimal Dispatch Based on
  Deep Reinforcement Learning.}
\newblock \textit{2021 IEEE Sustainable Power and Energy Conference (iSPEC)}:
  pp. 1082--1086. 2021.
\newblock \doi{10.1109/iSPEC53008.2021.9735756}.

\bibitem{a258_LIANG2024123390}
Liang, Tao, Chai, Lulu, Tan, Jianxin, Jing, Yanwei and Lv, Liangnian.
\newblock \enquote{Dynamic optimization of an integrated energy system with
  carbon capture and power-to-gas interconnection: A deep reinforcement
  learning-based scheduling strategy.}
\newblock \textit{Applied Energy} Vol. 367 (2024): p. 123390.
\newblock \doi{https://doi.org/10.1016/j.apenergy.2024.123390}.

\bibitem{a227_wang}
Wang, Ying, Zhou, Zhi, Botterud, Audun, Zhang, Kaifeng and Ding, Qia.
\newblock \enquote{Stochastic coordinated operation of wind and battery energy
  storage system considering battery degradation.}
\newblock \textit{Journal of Modern Power Systems and Clean Energy} Vol.~4
  No.~4 (2016): pp. 581--592.
\newblock \doi{10.1007/s40565-016-0238-z}.

\bibitem{a228_he}
He, Guannan, Chen, Qixin, Kang, Chongqing, Pinson, Pierre and Xia, Qing.
\newblock \enquote{Optimal Bidding Strategy of Battery Storage in Power Markets
  Considering Performance-Based Regulation and Battery Cycle Life.}
\newblock \textit{IEEE Transactions on Smart Grid} Vol.~7 No.~5 (2016): pp.
  2359--2367.
\newblock \doi{10.1109/TSG.2015.2424314}.

\bibitem{sage2023gt}
Sage, Manuel, Staniszewski, Martin and Zhao, Yaoyao~Fiona.
\newblock \enquote{Optimal Economic Gas Turbine Dispatch with Deep
  Reinforcement Learning.}
\newblock \textit{IFAC-PapersOnLine} Vol.~56 No.~2 (2023): pp. 10039--10044.
\newblock \doi{https://doi.org/10.1016/j.ifacol.2023.10.871}.
\newblock 22nd IFAC World Congress.

\bibitem{sage2025_b_and_g_preprint}
Sage, Manuel, Al~Handawi, Khalil and Zhao, Yaoyao~Fiona.
\newblock \enquote{Deep Reinforcement Learning for Joint Dispatch of Battery
  Energy Storage Systems and Gas Turbines in Microgrids with Renewable Energy.}
  (2025).
\newblock Manuscript under review.

\bibitem{a66_sutton2018reinforcement}
Sutton, Richard~S and Barto, Andrew~G.
\newblock \textit{Reinforcement learning: An introduction}.
\newblock MIT press (2018).

\bibitem{a79_mnih2015human}
Mnih, Volodymyr, Kavukcuoglu, Koray, Silver, David, Rusu, Andrei~A, Veness,
  Joel, Bellemare, Marc~G, Graves, Alex, Riedmiller, Martin, Fidjeland,
  Andreas~K, Ostrovski, Georg et~al.
\newblock \enquote{Human-level control through deep reinforcement learning.}
\newblock \textit{nature} Vol. 518 No. 7540 (2015): pp. 529--533.
\newblock \doi{10.1038/nature14236}.

\bibitem{a121_schulmanPPO}
Schulman, John, Wolski, Filip, Dhariwal, Prafulla, Radford, Alec and Klimov,
  Oleg.
\newblock \enquote{Proximal policy optimization algorithms.}
\newblock \textit{arXiv preprint arXiv:1707.06347}  (2017).

\bibitem{aeso}
{Alberta Electric System Operator}.
\newblock \enquote{Market and system reporting.} (2023).
\newblock \url{https://www.aeso.ca/market/market-and-system-reporting/},
  visisted on 2023-05-27.

\bibitem{a206_STAFFELL2016_wind}
Staffell, Iain and Pfenninger, Stefan.
\newblock \enquote{Using bias-corrected reanalysis to simulate current and
  future wind power output.}
\newblock \textit{Energy} Vol. 114 (2016): pp. 1224--1239.
\newblock \doi{10.1016/j.energy.2016.08.068}.

\bibitem{w28_SEelyzer}
{Siemens Energy}.
\newblock \enquote{Hydrogen and Power-to-X solutions - Elyzer P-300 - Technical
  Data.} (2024).
\newblock
  \url{https://www.siemens-energy.com/global/en/home/products-services/product-offerings/hydrogen-solutions.html},
  visisted on 2024-12-12.

\bibitem{a256_GHAIB2018433}
Ghaib, Karim and Ben-Fares, Fatima-Zahrae.
\newblock \enquote{Power-to-Methane: A state-of-the-art review.}
\newblock \textit{Renewable and Sustainable Energy Reviews} Vol.~81 (2018): pp.
  433--446.
\newblock \doi{https://doi.org/10.1016/j.rser.2017.08.004}.

\bibitem{a257_ALBREIKI}
Al-Breiki, Mohammed and Bicer, Yusuf.
\newblock \enquote{Techno-economic evaluation of a power-to-methane plant :
  Levelized cost of methane, financial performance metrics, and sensitivity
  analysis.}
\newblock \textit{Chemical Engineering Journal} Vol. 471 (2023): p. 144725.
\newblock \doi{https://doi.org/10.1016/j.cej.2023.144725}.

\bibitem{w27_co2cost}
IEA.
\newblock \enquote{Putting {CO2} to Use.}
\newblock \textit{International Energy Agency}
  (2019)\urlprefix\url{https://www.iea.org/reports/putting-co2-to-use}.

\bibitem{gymnasium}
Towers, Mark, Kwiatkowski, Ariel, Terry, Jordan, Balis, John~U., Cola,
  Gianluca~De, Deleu, Tristan, Goulão, Manuel, Kallinteris, Andreas, Krimmel,
  Markus, KG, Arjun, Perez-Vicente, Rodrigo, Pierré, Andrea, Schulhoff,
  Sander, Tai, Jun~Jet, Tan, Hannah and Younis, Omar~G.
\newblock \enquote{Gymnasium: A Standard Interface for Reinforcement Learning
  Environments.} (2024).
\newblock \eprint{2407.17032}.

\bibitem{stable-baslines3}
Raffin, Antonin, Hill, Ashley, Gleave, Adam, Kanervisto, Anssi, Ernestus,
  Maximilian and Dormann, Noah.
\newblock \enquote{Stable-Baselines3: Reliable Reinforcement Learning
  Implementations.}
\newblock \textit{Journal of Machine Learning Research} Vol.~22 No. 268 (2021):
  pp. 1--8.
\newblock \urlprefix\url{http://jmlr.org/papers/v22/20-1364.html}.

\bibitem{optuna}
Akiba, Takuya, Sano, Shotaro, Yanase, Toshihiko, Ohta, Takeru and Koyama,
  Masanori.
\newblock \enquote{Optuna: A next-generation hyperparameter optimization
  framework.}
\newblock \textit{Proceedings of the 25th ACM SIGKDD international conference
  on knowledge discovery \& data mining}: pp. 2623--2631. 2019.
\newblock \doi{10.1145/3292500.3330701}.

\bibitem{cem}
Szita, István and Lörincz, András.
\newblock \enquote{Learning Tetris Using the Noisy Cross-Entropy Method.}
\newblock \textit{Neural Computation} Vol.~18 No.~12 (2006): pp. 2936--2941.
\newblock \doi{10.1162/neco.2006.18.12.2936}.

\bibitem{gurobi}
{Gurobi Optimization, LLC}.
\newblock \enquote{{Gurobi Optimizer Reference Manual}.} (2024).
\newblock \urlprefix\url{https://www.gurobi.com}.

\bibitem{a229_tran_bes}
Tran, Duong and Khambadkone, Ashwin~M.
\newblock \enquote{Energy Management for Lifetime Extension of Energy Storage
  System in Micro-Grid Applications.}
\newblock \textit{IEEE Transactions on Smart Grid} Vol.~4 No.~3 (2013): pp.
  1289--1296.
\newblock \doi{10.1109/TSG.2013.2272835}.

\bibitem{a230_cheng_bes}
Cheng, Yu-Shan, Liu, Yi-Hua, Hesse, Holger~C., Naumann, Maik, Truong, Cong~Nam
  and Jossen, Andreas.
\newblock \enquote{A PSO-Optimized Fuzzy Logic Control-Based Charging Method
  for Individual Household Battery Storage Systems within a Community.}
\newblock \textit{Energies} Vol.~11 No.~2 (2018).
\newblock \doi{10.3390/en11020469}.

\bibitem{w23_cole_cost}
Cole, Wesley and Karmakar, Akash.
\newblock \enquote{Cost Projections for Utility-Scale Battery Storage: 2023
  Update.}
\newblock \textit{National Renewable Energy Lab (NREL), Golden, CO (United
  States)}  (2023)\doi{10.2172/1984976}.

\bibitem{w18_turbomachinery}
Robb, Drew.
\newblock \enquote{Aeroderivative gas turbines.}
  (2017)\url{https://www.turbomachinerymag.com/view/aeroderivative-gas-turbines},
  visited on 2022-11-25.

\bibitem{w17_energy2020capital}
{U.S. Energy Information Administration}.
\newblock \enquote{Capital Cost and Performance Characteristic Estimates for
  Utility Scale Electric Power Generating Technologies.}
  (2020)\urlprefix\url{https://www.eia.gov/analysis/studies/powerplants/capitalcost/pdf/capital_cost_AEO2020.pdf}.

\end{thebibliography}

\end{document}